\documentstyle[preprint,aps,epsf,floats,tighten]{revtex} 

\begin{document}

\preprint{\tighten \vbox{\hbox{hep-ph/9911396}
\hbox{} \hbox{} \hbox{} \hbox{} \hbox{} } }

\title{Determining $|V_{ub}|$ from the sum rule for semileptonic $B$ decay}

\author{Changhao Jin}

\address{School of Physics,
University of Melbourne\\Victoria 3010, Australia}

\maketitle

{\tighten
\begin{abstract}%
A precise determination of $|V_{ub}|$ can be obtained exploiting 
the sum rule for inclusive charmless semileptonic $B$-meson decays. The sum 
rule is derived on the basis of light-cone expansion and $b$-flavored 
quantum number 
conservation. The sum rule does not receive any perturbative QCD correction. 
In this determination of $|V_{ub}|$, there is no perturbative QCD 
uncertainty, while the dominant hadronic uncertainty is avoided. Moreover, 
this method is not only theoretically quite clean, but experimentally also
very efficient in the discrimination between $B\to X_u\ell\nu$ signal and 
$B\to X_c\ell\nu$ background. The sum rule requires measuring the lepton pair 
spectrum. We analyze the lepton pair spectrum, including the 
leading perturbative and nonperturbative QCD corrections. 
\end{abstract}
}

\newpage

\section{Introduction}
The measurement of the Cabibbo-Kobayashi-Maskawa (CKM) matrix \cite{ckm} 
element $|V_{ub}|$ is currently one of the important goals of $B$ physics.
Standard model predictions employ the fundamental parameter $|V_{ub}|$ as
input.  As the experiments at the $B$ factories are starting to
take data, a precise determination of $|V_{ub}|$ is increasingly vital for 
testing the standard model. For example, the standard model prediction for 
the CP-violating
asymmetry in $B\to J/\psi K_S$ decays depends on the value of $|V_{ub}|$.
Evidence for the CP-violating asymmetry has already been provided by the 
CDF Collaboration \cite{cdf}. As the dedicated experiments are expected to 
achieve much greater precision, an increase in the accuracy of $|V_{ub}|$ is 
highly desirable in order to test whether the complex phase of the CKM matrix 
is the only source of CP violation. 
On the other hand, despite empirical successes,
the standard model is definitely not the final theory of
particle physics. There remain many outstanding issues: electroweak symmetry
breaking, fermion masses and mixing, CP violation, replication of 
families, etc. A precise determination of $|V_{ub}|$ is also an important
link in pursuit of deeper principles to extend the standard model.

$|V_{ub}|$ can be determined from both inclusive and exclusive charmless 
semileptonic $B$ meson decays. The precision of the determination is limited by
experimental and theoretical difficulties. Charmless semileptonic $B$ decays
are a rare process. The main experimental difficulty in observing signals
from $b\to u\ell\nu$ processes is the very large background due to
$b\to c\ell\nu$ whose branching fraction is two orders of magnitude 
larger. On the theoretical side, calculations are needed in order to 
relate the measured quantity to $|V_{ub}|$ to extract a value from the data. 
Strong interaction
effects complicate the calculations, causing theoretical uncertainties.
The uncertainties result not just from nonperturbative QCD which is
notoriously difficult to calculate, but also from
perturbative QCD although in priciple it is calculable in perturbation 
theory. With the developments in nonperturbative techniques, 
the uncertainty from perturbative QCD due to the truncation of perturbation
series in practical calculations can be comparable to that from 
nonperturbative QCD, indicating that these two sources of uncertainty can be 
equally important.

Two approaches have been developed for a QCD treatment of inclusive decays of
heavy hadrons. One called the heavy quark expansion 
approach \cite{hqe} is based on the operator product expansion. 
One called the light-cone
approach \cite{jp,jin1,jp1,jin2,jin} is based on the light-cone expansion,
using the notion and method of deep inelastic lepton-hadron scattering.
In both approaches the heavy quark effective theory \cite{hqet} is exploited
as a limit of QCD.
There is important distinction between the two approaches. The former
is a short-distance expansion in local operators of increasing 
dimension, while the latter is a non-local light-cone expansion in 
matrix elements of increasing twist. Moreover,
the heavy quark expansion approach invokes quark-hadron duality, which
assumes rates evaluated at the parton level to be equal to observable rates
summed over a sufficient number of hadronic channels. Quark-hadron duality
cannot be exact \cite{dual}. The heavy quark expansion approach has to 
use quark kinematics and cannot account for the rate due to the extension of 
phase space from the parton level to the hadron level. 
The rate missing \cite{contri} is
just a manifestation of duality violation. In contrast, the light-cone
approach does not rely on the assumption of quark-hadron duality. Rates are
evaluated in the light-cone approach using physical phase space at the hadron
level. The light-cone approach was criticized in Refs.~\cite{ural}.
However, it has been shown \cite{contri} that that criticism is mistaken. 
The validity of the light-cone approach has been 
tested \cite{jp1} experimentally on the lepton energy spectrum in 
inclusive semileptonic decays of $B$ mesons. It is important to test
theoretical approaches experimentally in various ways.
 
To overcome the aforementioned difficulties in the determination of $|V_{ub}|$,
we proposed \cite{new} a method of extracting 
$|V_{ub}|$ from inclusive charmless semileptonic decays of $B$ mesons.
We proposed to measure the lepton pair spectrum, i.e., the decay distribution 
of the kinematic variable $\xi_u=(q^0+|{\bf q}|)/M_B$
in the $B$-meson rest frame, where $q$ is the momentum transfer to the lepton 
pair and $M_B$ denotes the $B$ meson mass.
Because most of $B\to X_u\ell\nu$ ($\ell=e$ or $\mu$) events have a value of 
$\xi_u$ beyond the threshold allowed for $B\to X_c\ell\nu$ decay, 
$\xi_u > 1-M_D/M_B = 0.65$ with $M_D$ being the $D$ meson mass, this
kinematic requirement provides a powerful tool for background 
suppression. If nothing else, the decay distribution of $\xi_u$ is of direct 
interest from an experimental standpoint. The sum rule can then be used to 
extract $|V_{ub}|$ from the
weighted integral of the measured $\xi_u$ spectrum. The sum rule 
follows from the light-cone expansion for inclusive charmless 
semileptonic decays of $B$ mesons and $b$-flavored quantum number conservation.
The sum rule is thus independent of phenomenological models. Moreover, the sum 
rule does not receive any perturbative QCD correction. Therefore, this 
method is not only experimentally very efficient, but theoretically also 
quite clean, allowing
a precise determination of $|V_{ub}|$ with a minimal overall 
(experimental and theoretical) error. 

The sum rule requires measuring the $\xi_u$ spectrum in inclusive charmless
semileptonic $B$ meson decays, weighted with $\xi_u^{-5}$.  
The purpose of the present paper is to refine the analysis of \cite{new}. We 
calculate the perturbative QCD correction to the $\xi_u$ spectrum to order
$\alpha_s$. We investigate in detail how gluon radiation and hadronic 
bound state effects affect the shape of the $\xi_u$ spectrum. 

The rest of the paper is organized as
follows. In Sec.~II the sum rule for inclusive charmless semileptonic
decays of $B$ mesons is briefly described. We discuss how to extract $|V_{ub}|$
from experiment exploiting the sum rule. In Sec.~III we study the $\xi_u$ 
spectrum,
including both the leading perturbative and nonperturbative QCD corrections. 
The exploration is extended to the weighted $\xi_u$ spectrum in Sec.~IV.
Finally in Sec.~V we present our conclusions.

\section{Sum Rule}
The sum rule for the decay distribution of $\xi_u$
in inclusive charmless semileptonic decays of
$B$ mesons has been derived \cite{new} from a nonperturbative treatment of 
QCD based on light-cone expansion.
The large $B$ meson mass sets a hard energy scale in the reactions, so that
the light-cone expansion is applicable \cite{jp,jin1,jp1,jin2,jin}
to inclusive $B$ decays that are dominated by light-cone singularities, just
as deep inelastic scattering.
For inclusive charmless semileptonic decays of $B$ mesons, $96\%$ of phase
space has $q^2\geq 1$ GeV$^2$. This dominance of the high-$q^2$ region in the
whole phase space $0\leq q^2\leq M_B^2$ renders the light-cone expansion 
especially reasonable. 
The theoretical expression for the weighted integral of the distribution of
$\xi_u$ 
\begin{equation}
S\equiv\int_0^1 d\xi_u\, \frac{1}{\xi_u^5}
\frac{d\Gamma}{d\xi_u}(B\to X_u\ell\nu) 
\label{eq:integral}
\end{equation}
is very clean. In the leading-twist approximation of QCD, 
it was found \cite{new} that
\begin{equation}
S = |V_{ub}|^2\frac{G_F^2M_B^5}{192\pi^3}
\langle B|\bar{b}\gamma^\mu b|B\rangle
\frac{P_\mu}{2M_B^2} \,\, ,
\label{eq:leading}
\end{equation}
where $P$ is the momentum of the decaying $B$ meson. 
The masses of leptons, the $u$ quark and the $\pi$ meson are neglected.
Because $b$-flavored quantum number conservation under strong interactions 
implies
\begin{equation}
\langle B|\bar{b}\gamma^\mu b|B\rangle = 2P^\mu ,
\label{eq:bnumber}
\end{equation}
the leading-twist contribution of nonperturbative QCD to the observable $S$ in
Eq.~(\ref{eq:leading}) is precisely calculable from first
principles. The result is the sum rule \cite{new}
\begin{equation}
S\equiv\int_0^1 d\xi_u\, \frac{1}{\xi_u^5}
\frac{d\Gamma}{d\xi_u}(B\to X_u\ell\nu) 
= |V_{ub}|^2\frac{G_F^2M_B^5}{192\pi^3}\, .
\label{eq:sumrule}
\end{equation}

The conserved current $b\gamma^\mu b$ is not subjected to renormalization
and the corresponding anomalous dimension vanishes thanks
to the Ward-Takahashi identity. Hence we recognize that the sum rule
(\ref{eq:sumrule}) does not receive any perturbative QCD correction. 

It might seem surprising that one is able to calculate $S$ so precisely in
QCD when the experimentalist measures hadrons, and it is well known that
hadrons cannot be understood completely within perturbation theory. There is
a physical way to make clear that we really do not need to know anything
about how hadrons are formed in order to predict $S$ in inclusive $B$ decays.
The observable $S$ measures something fundamental -- namely
the $b$-flavored quantum number carried by the $B$ meson, and
is basically determined by the underlying global symmetry of QCD. It is
therefore insensitive to hadronic bound state effects.

A measurement of the observable quantity $S$ in inclusive charmless 
semileptonic decays of $B$ mesons will lead to a
determination of $|V_{ub}|$ by 
use of the sum rule (\ref{eq:sumrule}) which contains no additional parameter. 
This determination of $|V_{ub}|$ is free of perturbative QCD 
uncertainties. Meanwhile, the dominant hadronic uncertainty is avoided. 
Higher-twist corrections of nonperturbative QCD to the sum rule 
(\ref{eq:sumrule}) are suppressed by a power of $\Lambda^2_{\rm QCD}/M_B^2$. 
Once the observable $S$ is measured, then Eq.~(\ref{eq:sumrule}) yields
$|V_{ub}|$ with small theoretical uncertainty.  Moreover, 
this method is not only theoretically quite clean, but experimentally also 
very efficient in the discrimination between $b\to u$ signal and $b\to c$
background, as we shall discuss. 

\section{The $\xi_{\lowercase{u}}$ spectrum}
In this section we describe the theoretical 
prediction for the $\xi_u$ spectrum. We explore step by step perturbative and
nonperturbative QCD effects on the $\xi_u$ spectrum.
We take into account first the perturbative
QCD correction to order $\alpha_s$ and then the leading nonperturbative QCD
correction.

\subsection{Leading perturbative QCD correction}
Ignoring QCD corrections, the tree-level $\xi_u$ spectrum in the 
free quark decay $b\to u\ell\nu$ in the $b$-quark rest frame takes the form
\begin{equation}
\frac{1}{\Gamma_0}\frac{d\Gamma_0}{d\xi_u}(b\to u\ell\nu) = 
\delta\left(\xi_u-\frac{m_b}{M_B}\right) ,
\label{eq:tree}
\end{equation}
with
\begin{equation}
\Gamma_0=\frac{G_F^2 m_b^5|V_{ub}|^2}{192\pi^3}\, ,
\label{eq:ga0}
\end{equation}
where $m_b$ represents the $b$-quark mass.
The resulting spectrum is a discrete line at $\xi_u= m_b/M_B$. This is 
simply a consequence of kinematics that fixes $\xi_u$ to the single value
$m_b/M_B$, no other values of $\xi_u$ are kinematically allowed in  
$b\to u\ell\nu$ decays. This is also the case for $b\to u\ell\nu$ processes 
with virtual gluon
emission. Therefore, the $\xi_u$ spectrum in free quark decays is still
a discrete line at $\xi_u= m_b/M_B$, shown in Fig.~1, 
even if virtual gluon emission occurs. However, the above kinematic 
relation no longer
holds for free $b$-quark decays with gluon bremsstrahlung, and hence
the spectrum expands downward below the parton-level end point  
$\xi_u= m_b/M_B$.

\begin{figure}[t]
\centerline{\epsfysize=9truecm \epsfbox{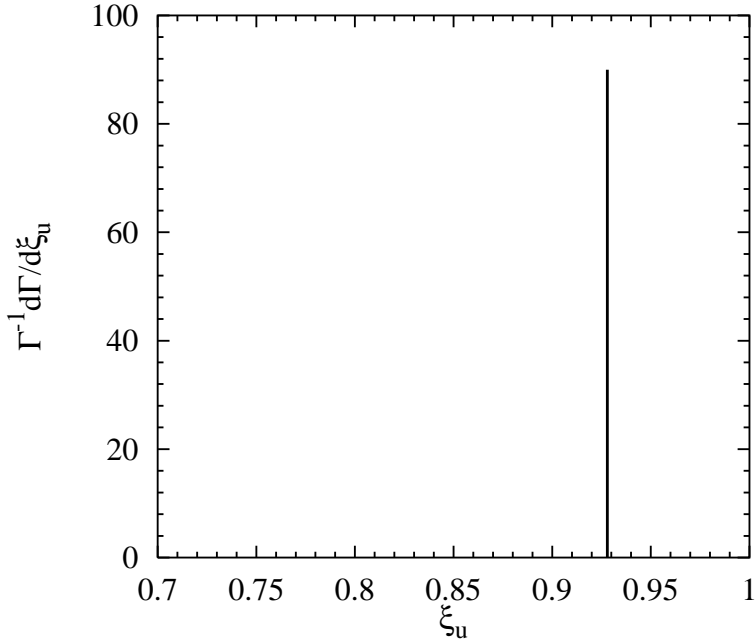}}
\tighten{
\caption[tau1]{The $\xi_u$ spectrum in the tree-level and virtual gluon
exchange processes $b\to u\ell\nu$. The $b$-quark mass $m_b=4.9$ GeV is used.} 
\label{fig:tree} }
\end{figure}

Including both virtual gluon emission and gluon bremsstrahlung, the 
differential decay rate as a function of $\xi_u$
in the $b$-quark rest frame is calculated to order $\alpha_s$ to be
\begin{eqnarray}
\frac{1}{\Gamma_0}\frac{d\Gamma}{d\xi_u}(b\to u\ell\nu) &=& \delta\left( \xi_u-
\frac{m_b}{M_B}\right)
\left[ 1-\frac{2\alpha_s}{3\pi}\left( \pi^2+\frac{13}{72}\right)\right] 
\nonumber\\
\nonumber\\
&+& \frac{2\alpha_s}{3\pi}\frac{M_B}{m_b}
\Bigg[ -2\left(\frac{{\rm ln}r}{r}\right)_+ 
-\frac{13}{3}\left(\frac{1}{r}\right)_+ 
\nonumber\\
\nonumber\\
&+& \frac{79}{9}+\frac{407}{36}r-\frac{367}{12}r^2+\frac{59}{3}r^3
-\frac{50}{9}r^4
+\frac{11}{12}r^5-\frac{7}{36}r^6 
\nonumber\\
\nonumber\\
&+& \left( -\frac{2}{3}+\frac{23}{3}r+3r^2-\frac{8}{3}r^3\right) {\rm ln}r
+2r^2(-3+2r){\rm ln}^2 r \Bigg] ,
\label{eq:free}
\end{eqnarray}
where $r=1-M_B\xi_u/m_b$ varying in the range $0\leq r\leq 1$, 
corresponding to the kinematic
range $0\leq\xi_u\leq m_b/M_B$ at the parton level of quarks and gluons. 
The distribution
$[g(r)]_+$ is defined to coincide with the function $g(r)$ for all values of 
$r$ 
greater than 0, and to have a singularity at $r=0$ such that the integral of
this distribution with any smooth function $t(r)$ gives
\begin{equation}
\int_0^1 dr\, [g(r)]_+ t(r)
=\int_0^1 dr\, g(r)[t(r)-t(0)] .
\label{eq:def1}
\end{equation}
The result for the perturbative $\xi_u$ spectrum with the QCD radiative 
correction\footnote{The sum rule (\ref{eq:sumrule}) does not receive any 
perturbative QCD correction
because it results from the conserved current. The numerical calculation
by integrating Eq.~(\ref{eq:free}) with the factor $\xi_u^{-5}$ has not 
reached the absolute zero but yielded a tiny correction (about $0.5\%$). 
We have checked that this is 
only a numerical artifact due to the limited accuracy of numerical integration
by computer, especially given the rapid oscillation of the integrand near the 
end point $\xi_u=0$.} 
to order $\alpha_s$ is shown in Fig.~2. 
The perturbative $\xi_u$ spectrum is singular at its end point $\xi_u=m_b/M_B$
due to infrared divergences. We observe that gluon bremsstrahlung generates a 
small tail below $\xi_u=m_b/M_B$.

\begin{figure}[t]
\centerline{\epsfysize=9truecm \epsfbox{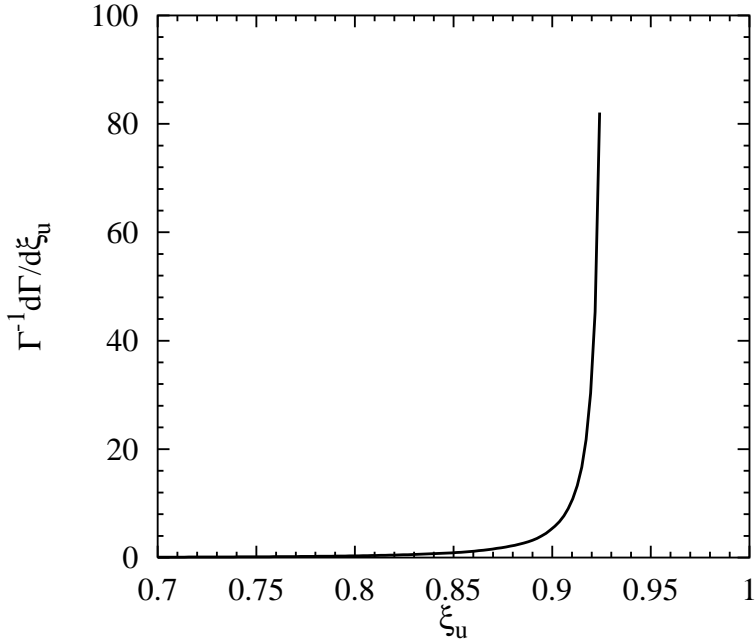}}
\tighten{
\caption[tau1]{The perturbative $\xi_u$ spectrum with the QCD radiative
correction to order $\alpha_s$, obtained from input values $m_b=4.9$ GeV and 
$\alpha_s=0.2$.} 
\label{fig:pert} }
\end{figure}

Integrating Eq.~(\ref{eq:free}) over $\xi_u$ yields the total perturbative
decay rate
\begin{equation}
\Gamma (b\to u\ell\nu)= \Gamma_0\left[1-\frac{2\alpha_s}{3\pi}\left(\pi^2-
\frac{25}{4}\right)\right] .
\label{eq:total}
\end{equation}
This agrees with the well-known result obtained in \cite{tot}.

\subsection{Leading nonperturbative QCD correction}
To calculate the real physical decay distribution in inclusive decays
$B\to X_u\ell\nu$, perturbative QCD alone is not sufficient.  We must
also account for hadronic bound state effects due to the confinement of the 
$b$ quark inside the $B$ meson. 
We now consider nonperturbative QCD effects on the $\xi_u$ spectrum.
In the framework of the light-cone expansion, the leading nonperturbative
QCD effect is incorporated in the $b$-quark distribution 
function \cite{jin}
\begin{equation}
f(\xi) = \frac{1}{4\pi}\int \frac{d(y\cdot P)}{y\cdot P}\, e^{i\xi y\cdot P}
\langle B|\bar{b}(0)y\!\!\!/{\cal P}exp[ig_s\int_y^0 dz^\mu A_\mu (z)]b(y)
|B\rangle |_{y^2=0}\, ,
\label{eq:def}
\end{equation}
where ${\cal P}$ denotes path ordering.
The distribution function $f(\xi)$ has a simple physical interpretation:
It is the probability of finding a $b$-quark with momentum $\xi P$ inside 
the $B$ meson with momentum $P$. The real physical spectrum is then obtained 
from a convolution of 
the hard perturbative spectrum with the soft nonperturbative distribution 
function:
\begin{equation}
\frac{d\Gamma}{d\xi_u}(B\to X_u\ell\nu) = \int_{\xi_u}^1 d\xi\, f(\xi)
\frac{d\Gamma}{d\xi_u}(b\to u\ell\nu, p_b=\xi P) ,
\label{eq:convol}
\end{equation}
where the $b$-quark momentum $p_b$ in the perturbative spectrum is replaced by
$\xi P$. The analytic result for the perturbative 
spectrum to order $\alpha_s$ is given in
Eq.~(\ref{eq:free}). The interplay between nonperturbative and perturbative
QCD effects has been accounted for since confinement implies that free quarks 
are not asymptotic states of the theory and the separation of perturbative and
nonperturbative effects cannot be done in a clear-cut way.

Equation (\ref{eq:convol}) demonstrates that the physical $\xi_u$ spectrum
depends on the distribution function $f(\xi)$. The definition of it, 
Eq.~(\ref{eq:def}), involving the $B$-meson matrix element of the non-local
$b$-quark operators separated along the light cone,
makes clear that the distribution function is a 
nonperturbative quantity.  
Although a complete calculation of the distribution function in QCD is
impossible at present due to our ignorance of nonperturbative QCD,
some basic properties of it are known \cite{jin}.
The distribution function is universal in
the sense that the same distribution function also summarizes the leading
nonperturbative QCD contribution in inclusive radiative $B$ decays
$B\to X_s\gamma$. 
It is gauge invariant and obeys positivity. It has a support between 0 and 1
and is exactly normalized to unity because of $b$-flavored quantum number 
conservation\footnote{The same conservation law results in the sum rule 
(\ref{eq:sumrule}).}. It contains the free quark decay as a limiting case with
$f(\xi)=\delta(\xi-m_b/M_B)$. In the free quark limit, Eq.~(\ref{eq:convol}) 
consistently reproduces the free quark spectrum. 

In addition, the mean $\mu$ and the variance
$\sigma^2$ of the distribution function were deduced \cite{jin} 
using operator product expansion and heavy quark effective 
theory (HQET) \cite{hqet}:
\begin{equation}
\mu\equiv \int_0^1 d\xi\, \xi f(\xi)= \frac{m_b}{M_B}
\left( 1+\frac{5E_b}{3}\right) ,
\label{eq:mean}
\end{equation}
\begin{equation}
\sigma^2\equiv \int_0^1 d\xi\, (\xi-\mu)^2 f(\xi)= 
\left(\frac{m_b}{M_B}\right)^2
\left[\frac{2K_b}{3}-\left(\frac{5E_b}{3}\right)^2\right] ,
\label{eq:variance}
\end{equation}
where $E_b = K_b+G_b$ and $K_b$ and $G_b$ are the dimensionless HQET 
parameters of order $(\Lambda_{\rm QCD}/m_b)^2$, which are often referred to 
by the alternate names
$\lambda_1 = -2m_b^2K_b$ and $\lambda_2 = -2m_b^2G_b/3$. The parameter
$\lambda_2$ can be extracted from the $B^\ast-B$ mass splitting:
$\lambda_2=(M^2_{B^\ast}-M^2_B)/4=0.12$ GeV$^2$. The parameter $\lambda_1$
suffers from large uncertainty. 

The mean value and variance of the distribution function characterize the 
location of the ``center of mass'' of the distribution function and the
square of its width, respectively. They specify the primary shape of the
distribution function. From Eqs.~(\ref{eq:mean}) and (\ref{eq:variance}) we
know that the distribution function is sharply peaked around
$\xi = \mu \approx m_b/M_B$ close to 1 and its width of order 
$\Lambda_{\rm QCD}/M_B$ is narrow, suggesting that the distribution function
is close to the delta function form in the free quark limit.

Nonperturbative QCD methods such as lattice simulation 
and QCD sum rules could help determine further the functional 
form of the distribution 
function. The distribution function can also be extracted directly from
experiments of inclusive semileptonic \cite{new} or radiative \cite{jin}
decays of $B$ mesons. The universality of the distribution function implies
great predictive power: Once the distribution function is measured from one 
process, it can be used to make predictions
in all other processes in a model-independent manner. Since these are as yet
not done, we perform the calculations using the parametrization \cite{jin1}
of the distribution function
\begin{equation}
f(\xi)=N \frac{\xi(1-\xi)^\alpha}{[(\xi-a)^2+b^2]^\beta}\theta(\xi)
\theta(1-\xi) ,
\label{eq:para}
\end{equation}
where $\alpha$, $\beta$, $a$, and $b$ are four parameters and $N$ is the
normalization constant. The parametrization (\ref{eq:para}) respects all the 
known properties of the distribution function, in particular the strong
constraints of the sum rules (\ref{eq:mean}) and (\ref{eq:variance}).

The $\xi_u$ spectrum can be calculated using Eqs.~(\ref{eq:convol}),
(\ref{eq:free}) and (\ref{eq:para}). Including both the leading 
nonperturbative and
perturbative QCD corrections, we obtain the $\xi_u$ spectrum shown in
Fig.~3, using $\alpha_s=0.2$, $\alpha=\beta=1$, $a=0.9548$ and $b=0.005444$. 
Here both values of $\alpha$ and 
$\beta$ for the four-parameter distribution function (\ref{eq:para}) are 
preset to be 1, but in general they need not be integers. The values of $a$ 
and $b$ are then inferred from the 
sum rules (\ref{eq:mean}) and (\ref{eq:variance}) using $m_b= 4.9$ GeV and
$\lambda_1= -0.5$ GeV$^2$, giving the mean $\mu= 0.93$ and the variance
$\sigma^2= 0.006$ for the distribution function. 

Bound-state effects lead to 
the extension of phase space from the parton level to the hadron level, also
stretch the spectrum downward below $m_b/M_B$, and are solely responsible for
populating the spectrum upward in the gap between the parton-level end point
$\xi_u= m_b/M_B$ and the hadron-level end point $\xi_u= 1$. The interplay 
between nonperturbative and perturbative QCD effects eliminates the singularity
at the end point of the perturbative spectrum, so that the physical spectrum
shows a smooth behavior over the entire range of $\xi_u$, $0\leq\xi_u\leq 1$, 
as in Fig.~3. Therefore, nonperturbative QCD effects play a crucial role in
shaping the $\xi_u$ spectrum. 

\begin{figure}[t]
\centerline{\epsfysize=9truecm \epsfbox{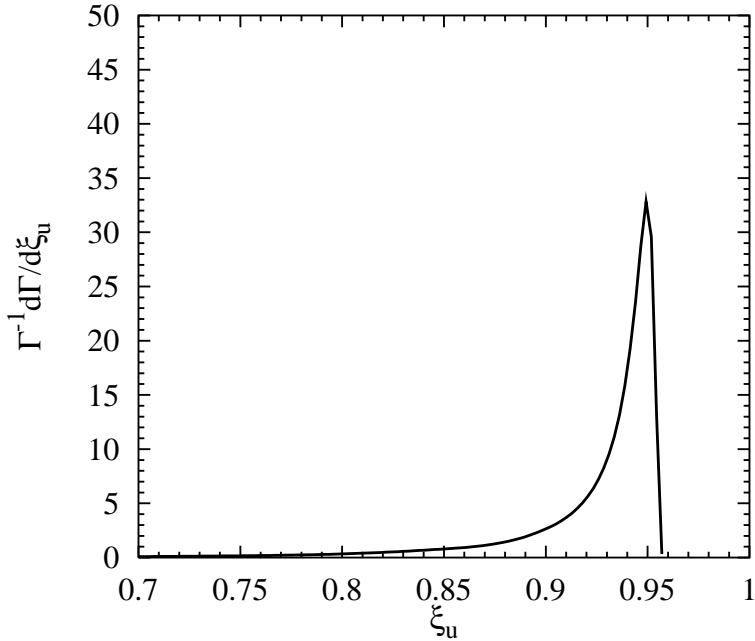}}
\tighten{
\caption[tau1]{The $\xi_u$ spectrum including both the leading nonperturbative
and perturbative QCD corrections.} 
\label{fig:real} }
\end{figure}

By integrating Eq.~(\ref{eq:convol}) over $\xi_u$, we obtain the fraction of 
$B\to X_u\ell\nu$ events above the charm threshold 
allowed for the predominant $B\to X_c\ell\nu$ decays, defined as
\begin{equation}
R\equiv\frac{1}{\Gamma(B\to X_u\ell\nu)}
\int_{1-M_D/M_B}^1 d\xi_u\, \frac{d\Gamma}{d\xi_u}(B\to X_u\ell\nu) .
\label{eq:fraction}
\end{equation}
We find that $R = 79\%$ for the spectrum shown in Fig.~3. This result refines
our previous result \cite{new}. The earlier calculation 
without QCD radiative corrections found $R = 99\%$.
Gluon bremsstrahlung streches the $\xi_u$ spectrum downward, and hence gives
rise to a decrease of the fraction of $B\to X_u\ell\nu$ events above 
the charm threshold.
Nevertheless, most of $B\to X_u\ell\nu$ events remain above the charm
threshold. Thus the kinematic cut on the observable quantity $\xi_u$ is very 
efficient in 
disentangling $B\to X_u\ell\nu$ signal from $B\to X_c\ell\nu$ background. 
This efficiency can be explained by the uniqueness of the $\xi_u$ spectrum:
The $\xi_u$ spectrum stemming from 
the quark-level and virtual gluon exchange processes would only concentrate
at $\xi_u= m_b/M_B$, shown in Fig.~1, solely on kinematic grounds, and gluon
bremsstrahlung and hadronic bound state effects smear the spectrum about this
point, but most of the decay rate remains at large values of $\xi_u$, as 
revealed by Figs.~2 and 3.

The fraction $R$ of course depends on forms of the distribution function.
However, we find that $R$ is relatively insensitive to forms of the
distribution function once the mean and variance of it, which
are known from HQET, given by Eqs.~(\ref{eq:mean}) and (\ref{eq:variance}),
are kept fixed. Therefore, the above calculation of $R$ can be considered as a 
typical estimate of the
fraction of $B\to X_u\ell\nu$ events above the charm threshold.

\section{The weighted $\xi_{\lowercase{u}}$ spectrum}
The weighted spectrum $\xi_u^{-5}d\Gamma(B\to X_u\ell\nu)/d\xi_u$ is more 
directly relevant for
the measurement of the observable $S$ in inclusive charmless semileptonic
decays of $B$ mesons. Measurements of $S$ can be obtained from an extrapolation
of the weighted spectrum measured above the charm threshold to the full phase
space available in $B\to X_u\ell\nu$ decays. 
While the normalization of the weighted spectrum given by the sum rule
(\ref{eq:sumrule}) does not
depend on the $b$-quark distribution function $f(\xi)$, thus being 
model-independent,
the shape of the weighted spectrum does. In this section we perform a detailed
analysis of the shape of the weighted spectrum, using Eqs.~(\ref{eq:convol}),
(\ref{eq:free}) and (\ref{eq:para}).

We first explore the impact of gluon radiation on the shape of the weighted
$\xi_u$ spectrum. We show in Fig.~4 the weighted $\xi_u$ spectram without
and with the QCD radiative correction. It is evident from Fig.~4 that the shape
of the weighted $\xi_u$ spectrum receives a significant correction due to
gluon radiation. However, the shape of the spectrum appears to be insensitive 
to the value 
of the strong coupling $\alpha_s$, varied within a reasonable range.
 
\begin{figure}[t]
\centerline{\epsfysize=9truecm \epsfbox{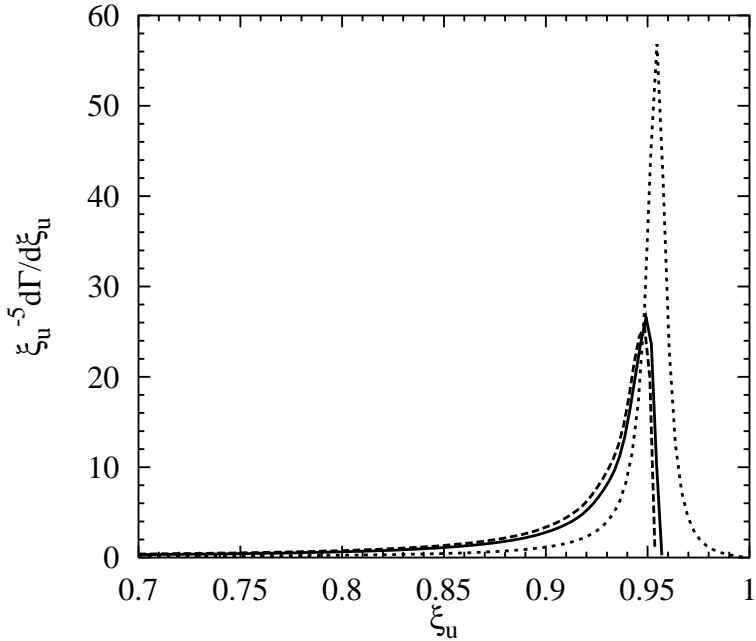}}
\tighten{
\caption[tau1]{The weighted $\xi_u$ spectrum without (dotted curve) and with 
(solid curve: $\alpha_s=0.2$, dashed curve: $\alpha_s=0.27$) the QCD radiative 
correction. The four parameters for the distribution function are taken to be
$\alpha=\beta=1$, $a=0.9548$, and $b=0.005444$. All the spectra are normalized
to have unit area.} 
\label{fig:radiation} }
\end{figure}

We investigate next the sensitivity of the spectrum to the form of the
distribution function, which would reflect the impact of hadronic bound state 
effects. For this purpose we choose to use two very different 
forms \cite{jin3} for the distribution 
function, albeit having the same mean value and variance. The calculated
weighted spectra are shown in Fig.~5, taking into account the QCD radiative  
correction to order $\alpha_s$. The weighted spectrum exhibits
a strong dependence on the form of the distribution function, 
even though the mean
value and variance of the distribution functions are the same. 

\begin{figure}[t]
\centerline{\epsfysize=9truecm \epsfbox{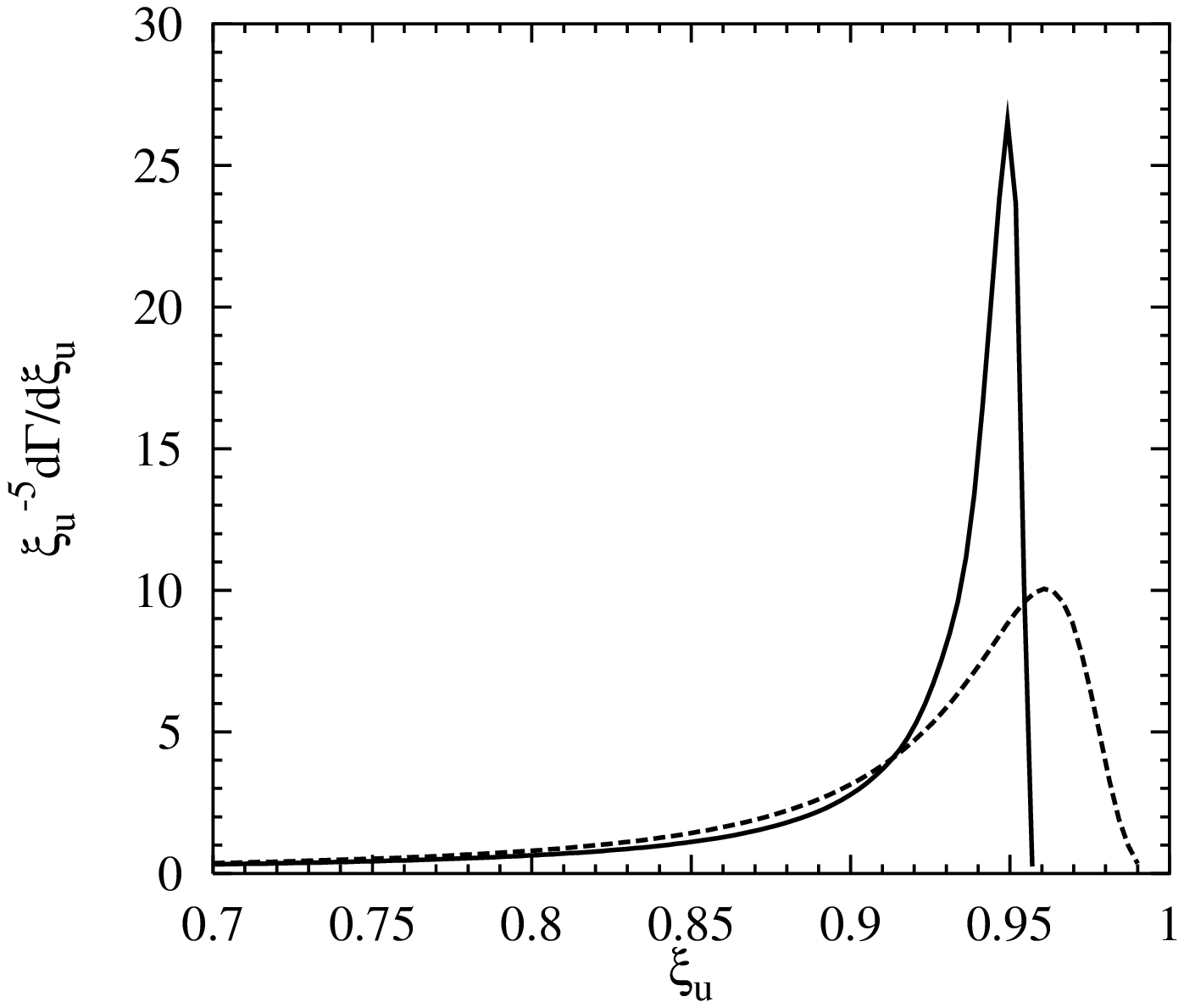}}
\tighten{
\caption[tau1]{Comparison of the weighted $\xi_u$ spectra in two forms of the
distribution function. The solid curve corresponds to the four-parameter
distribution function (\ref{eq:para}) with $\alpha=\beta=1$, $a=0.9548$, and 
$b=0.005444$, while the dashed curve corresponds to the four-parameter 
distribution
function (\ref{eq:para}) with $\alpha=\beta=2$, $a=0.9864$, and $b=0.02557$.
These two forms of the distribution function have the same mean value 
$\mu=0.93$ and 
variance $\sigma^2=0.006$. The strong coupling is taken to be $\alpha_s=0.2$. 
Both spectra are normalized to have unit area.} 
\label{fig:form} }
\end{figure}

Generally, since the quark-level processes, exclusive of gluon bremsstrahlung, 
generate a discrete line,
the shape of the $\xi_u$ spectrum directly reflects the inner 
long-distance dynamics of the reaction. This argument elucidates the strong
variation of the weighted spectrum with the form of the distribution function,
illustrated in Fig.~5.

In fact, ignoring QCD radiative 
corrections, one obtains from Eqs.~(\ref{eq:convol}) and (\ref{eq:tree}) the 
weighted spectrum in the $B$ rest frame
\begin{equation}
\frac{1}{\xi_u^5}\frac{d\Gamma}{d\xi_u}(B\to X_u\ell\nu) 
= \frac{G_F^2M_B^5|V_{ub}|^2}{192\pi^3} f(\xi_u) .
\label{eq:direct}
\end{equation}
It makes clear that the resulting
weighted spectrum is directly proportional to the distribution function.
In other words, the weighted $\xi_u$ spectrum is most sensitive 
to the distribution function.
This salient feature renders the weighted spectrum idealy suited for 
the direct
extraction of the distribution function from experiment. After an observed
spectrum $\xi_u^{-5}d\Gamma(B\to X_u\ell\nu)/d\xi_u$ is radiatively corrected, 
it is nothing but the $b$-quark distribution function if higher-twist terms
are neglected. 

Extrapolation of the weighted $\xi_u$ spectrum measured above the
charm threshold to the full $\xi_u$ range requires a theoretical calculation 
of the rate ratio
\begin{equation}
W\equiv \frac{1}{S}\int^1_{1-M_D/M_B} d\xi_u\, \frac{1}{\xi_u^5}
\frac{d\Gamma}{d\xi_u}(B\to X_u\ell\nu) ,
\label{eq:extrap}
\end{equation}
where $S$ is defined in Eq.~(\ref{eq:integral}).
We calculate the ratio $W$ using Eqs.~(\ref{eq:sumrule}), (\ref{eq:convol}),
(\ref{eq:free}) and (\ref{eq:para}), and
the theoretical uncertainties are estimated as follows:

We study the variation of $W$ with respect to the mean value and 
the variance
of the distribution function setting $\alpha=\beta=1$ in Eq.~(\ref{eq:para}).
Actually, this amounts
to the study of the ratio $W$ as functions of $m_b$ and $\lambda_1$,
since, essentially, the mean value of the distribution function is determined 
by the $b$-quark mass and its variance is determined by $\lambda_1$ according
to the sum rules in Eqs.~(\ref{eq:mean}) and (\ref{eq:variance}). 
At present, the estimated values of the $b$-quark mass and $\lambda_1$ vary 
in the ranges
\begin{eqnarray}
&&m_b=4.9\pm 0.15 \ {\rm GeV},\\
&&\lambda_1= -(0.5\pm 0.2) \ {\rm GeV}^2.
\end{eqnarray}
The variation of $m_b$ leads
to an uncertainty of $8\%$ in $W$ if other parameters are kept
fixed. A small uncertainty of $3\%$ in $W$ results from the
variation of $\lambda_1$. In other words, the ratio $W$ displays
a strong dependence on
the mean value of the distribution function of the $b$ quark 
inside the $B$ meson, but is insensitive to the variance of the 
distribution function.

We examine the further sensitivity of the rate ratio $W$ to the form of the 
distribution 
function when keeping the mean value and variance of it fixed, by varying 
the values of the two additional parameters $\alpha$ and $\beta$ in the
parametrization (\ref{eq:para}). We estimate that the variation of $W$ 
is $3\%$ if the form of the distribution function
is changed but with the same mean value and variance.

We estimate the uncertainty due to the truncation of the perturbative series
in Eq.~(\ref{eq:free})
by varying the renormalization scale between $m_b/2$ and $2m_b$. 
We find that an uncertainty of $8\%$ in the ratio $W$ 
stems from the renormalization scale dependence.

This analysis implies that at present the theoretical error in the calculation
of $W$ has two main sources: the value of $m_b$ (or equivalently, the mean
value of the distribution function) and the renormalization scale dependence.
Finally, adding all the uncertainties in quadrature we arrive at 
\begin{equation}
W = (76\pm 9)\% 
\end{equation}
with an error of $12\%$.

\section{Conclusions} 
We have studied the $\xi_u$ spectrum in inclusive charmless semileptonic decays
of $B$ mesons. The perturbative QCD correction to the spectrum is calculated to
order $\alpha_s$. The leading nonperturbative QCD effect is calculated using
light-cone expansion and heavy quark effective theory. The $\xi_u$ spectrum is
unique in that the tree-level and virtual gluon exchange processes 
$b\to u\ell\nu$ at the parton level generate a trivial $\xi_u$ spectrum -- a 
discrete line at
$\xi_u=m_b/M_B\approx 0.93$ in the $b$-quark rest frame, which is well above
the charm threshold $\xi_u=1-M_D/M_B=0.65$. Gluon bremsstrahlung results in
a small tail in the lepton pair spectrum below $m_b/M_B$. Bound-state effects 
lead 
to the extension of phase space from the parton level to the hadron level, 
also stretch the spectrum downward below $m_b/M_B$, and are solely responsible
for populating the spectrum upward in the gap between the parton-level 
endpoint $\xi_u=m_b/M_B$ and the hadron-level endpoint $\xi_u=1$, smoothing
out the singularity in the perturbative spectrum. 
As a result of these two distinct effects, the lepton pair spectrum spreads
over the entire physical range $0\leq\xi_u\leq 1$. Still, we find
that about $80\%$ of the lepton pair spectrum in $B\to X_u\ell\nu$ lies above 
the charm threshold,
$\xi_u> 1-M_D/M_B$. This kinematic cut is most efficient in the suppression of
the background of $B$ semileptonic decays into charmed particles.

The observable $S$ can be measured by the extrapolation of the weighted 
spectrum $\xi_u^{-5}d\Gamma(B\to X_u\ell\nu)/d\xi_u$ measured above the charm 
threshold to the entire phase space. Gluon
bremsstrahlung and hadronic bound state effects strongly affect the shape of
the weighted $\xi_u$ spectrum. However, the shape of the weighted $\xi_u$
spectrum is insensitive to the value of the strong coupling $\alpha_s$, varied
in a reasonable range. The overall picture appears to be that the weighted 
$\xi_u$ spectrum is peaked towards larger values of $\xi_u$ with a narrow 
width. The contribution below $\xi_u = 0.65$ is moderate and relatively 
insensitive to forms of the distribution function. This suggests that 
extrapolating the weighted $\xi_u$ spectrum down to low $\xi_u$ would not 
introduce a considerable uncertainty in the value of $S$.  Quantitatively,
our analysis determines the rate ratio for the extrapolation of the 
weighted $\xi_u$
spectrum to be $W=(76\pm 9)\%$ with an error of $12\%$ at present.

Another interesting use of the weighted $\xi_u$ spectrum is its utility for 
directly
extracting the $b$-quark distribution function. The universality of the
distribution function implies that the distribution function extracted from
inclusive semileptonic $B$ decays can be used to make predictions in inclusive
radiative decays $B\to X_s\gamma$ in a model-independent manner and vice versa.
In particular, a measurement of the distribution function from the photon
energy spectrum in $B\to X_s\gamma$ \cite{jin} would be very useful to 
improve the
measurement of $S$ from extrapolation of the weighted $\xi_u$ spectrum.  

An improved knowledge of the form of the distribution function is important
for an error reduction in extrapolation. More precisely,
if by direct measurements or theoretical studies the uncertainty in the mean 
value of the distribution function can be
improved from currently $3\%$ to $1.5\%$ in the future, the resulting
uncertainty in the ratio $W$ for the extrapolation would decrease from   
$8\%$ to $3\%$. A calculation of the $O(\alpha_s^2)$ perturbative QCD 
correction to the $\xi_u$ spectrum is also important for improving the
accuracy of the extrapolation.

The physical observable quantity $S$ is connected with $|V_{ub}|$ via the sum
rule (\ref{eq:sumrule}) for semileptonic $B$ decay in the leading-twist 
approximation of QCD. The sum rule is a fundamental, model-independent 
prediction of QCD. Note that the sum rule 
does not rely on the heavy quark effective theory. No arbitrary parameter other
than $|V_{ub}|$ enters the sum rule. The dominant hadronic 
uncertainty is avoided in 
the sum rule. As important, there is no perturbative QCD modification of this 
sum rule, so that the potential source of theoretical uncertainty 
associated with perturbative QCD calculations is totally averted. 

The kinematic cut on $\xi_u$, $\xi_u>1-M_D/M_B$, and the semileptonic $B$ 
decay sum rule, Eq.~(\ref{eq:sumrule}), offer an outstanding
opportunity for the precise determination of $|V_{ub}|$ from the observable
$S$. We wish to emphasize
that this method is both exceptionally clean theoretically and very 
efficient experimentally in background suppression. 

There remain two types of theoretical uncertainties in the  
determination of $|V_{ub}|$. First, higher-twist (or power suppressed) 
corrections to the sum rule cause a theoretical error of order 
$\Lambda^2_{\rm QCD}/M_B^2\sim 1\%$ on $|V_{ub}|$. A quantitative study of
higher-twist effects could further reduce this small theoretical uncertainty
associated with simply extracting the value of $|V_{ub}|$ from the measured
value of $S$.
Second, the extrapolation of the weighted $\xi_u$ spectrum to low $\xi_u$
gives rise to a systematic error for the measurement of $S$. The 
status of this additional theoretical uncertainty associated with 
extrapolation, as well as how to improve it, have been discussed above.

Eventually,
the error on $|V_{ub}|$ determined by this method would mainly depend
on how well the observable $S$ can be measured. To measure $S$ 
experimentally one needs to be able to reconstruct the neutrino. This poses
a challenge to experiment. Given the unique potential of determining 
$|V_{ub}|$, we would urge our experimental colleagues to examine the 
feasibility of the method.
  
\acknowledgments
This work is supported by the Australian Research Council.

{\tighten

} 


\begin{references}

\bibitem{ckm} N. Cabibbo, Phys. Rev. Lett. {\bf 10}, 531 (1963);\\
M. Kobayashi and T. Maskawa, Prog. Theor. Phys. {\bf 49}, 652 (1973). 

\bibitem{cdf} CDF Collaboration, T. Affolder {\it et al.}, 
Phys. Rev. D {\bf 61}, 072005 (2000).

\bibitem{hqe} J. Chay, H. Georgi, and B. Grinstein, Phys. Lett. B {\bf 247},
399 (1990);\\
I.I. Bigi, N.G. Uraltsev, and A.I. Vainshtein, Phys. Lett. B 
{\bf 293}, 430 (1992); {\bf 297}, 477(E) (1993);\\
I.I. Bigi, M.A. Shifman, N.G. Uraltsev, and A.I. Vainshtein, Phys.
Rev. Lett. {\bf 71}, 496 (1993); Int. J. Mod. Phys. A {\bf 9}, 2467 (1994);\\
A.V. Manohar and M.B. Wise, Phys. Rev. D {\bf 49}, 1310 (1994);\\
B. Blok, L. Koyrakh, M.A. Shifman, and A.I. Vainshtein, Phys. Rev. D {\bf 49},
3356 (1994); {\bf 50}, 3572(E) (1994);\\
M. Luke and M.J. Savage, Phys. Lett. B {\bf 321}, 88 (1994);\\
A.F. Falk, M. Luke, and M.J. Savage, Phys. Rev. D {\bf 49}, 3367 (1994);\\
T. Mannel, Nucl. Phys. B {\bf 413}, 396 (1994);\\
A.F. Falk, Z. Ligeti, M. Neubert, and Y. Nir, Phys. Lett. B {\bf 326}, 145
(1994);\\
M. Neubert, Phys. Rev. D {\bf 49}, 3392 and 4623 (1994);\\ 
T. Mannel and M. Neubert, Phys. Rev. D {\bf 50}, 2037 (1994).

\bibitem{jp} C.H. Jin and E.A. Paschos, in {\it Proceedings of the 
International 
Symposium on Heavy Flavor and Electroweak Theory}, Beijing, China, 
1995, edited by C.H. Chang and C.S. Huang (World Scientific, Singapore,
1996), p.~132; hep-ph/9504375.

\bibitem{jin1} C.H. Jin, Phys. Rev. D {\bf 56}, 2928 (1997).

\bibitem{jp1} C.H. Jin and E.A. Paschos, Eur. Phys. J. C {\bf 1}, 523 (1998).

\bibitem{jin2} C.H. Jin, Phys. Rev. D {\bf 56}, 7267 (1997).

\bibitem{jin} C.H. Jin, Eur. Phys. J. C {\bf 11}, 335 (1999).

\bibitem{hqet} N. Isgur and M.B. Wise, Phys. Lett. B {\bf 232}, 113 (1989);
{\bf 237}, 527 (1990);\\
E. Eichten and B. Hill, Phys. Lett. B {\bf 234}, 511 (1990);
{\bf 243}, 427 (1990);\\
H. Georgi, Phys. Lett. B {\bf 240}, 447 (1990).

\bibitem{dual} N. Isgur, Phys. Lett. B {\bf 448}, 111 (1999);\\
I.I. Bigi, hep-ph/0001003.

\bibitem{contri} C.H. Jin, hep-ph/9906212 v2.

\bibitem{ural} N. Uraltsev, Int. J. Mod. Phys. A {\bf 14}, 4641 (1999),
hep-ph/9905520;\\
I.I. Bigi, hep-ph/9907270.

\bibitem{new} C.H. Jin, Mod. Phys. Lett. A {\bf 14}, 1163 (1999).

\bibitem{tot} N. Cabibbo and L. Maiani, Phys. Lett. {\bf 79}B, 109 (1978).

\bibitem{jin3} C.H. Jin, Phys. Rev. D {\bf 57}, 6851 (1998).
\end{references}
\end{document}